\documentclass{ws-procs975x65}

\usepackage{amsmath}
\usepackage{amssymb}
\usepackage{amsfonts}

\begin{document}

\title{EXTENDED GRAVITY: STATE OF THE ART AND PERSPECTIVES}

\author{SALVATORE CAPOZZIELLO and MARIAFELICIA DE LAURENTIS}

\address{Dipartimento di Fisica, Università
di Napoli {}``Federico II'' and
INFN Sezione  di Napoli, \\Compl. Univ. di
Monte S. Angelo, Edificio G, Via Cinthia, I-80126, Napoli, Italy.\\
E-mail: capozziello@na.infn.it; felicia@na.infn.it\\
www.unina.it}

\begin{abstract}
Several issues coming from Cosmology, Astrophysics and Quantum Field Theory suggest to extend the  General Relativity in order to overcome  several shortcomings emerging at conceptual and experimental level. From one hand, standard Einstein theory fails as soon as one wants to achieve  a full quantum description of space-time. In fact, the lack of a final self-consistent Quantum Gravity Theory can be considered one of the starting points for  alternative theories of gravity. Specifically, the approach  based on corrections and enlargements of the Einstein scheme, have become a sort of paradigm in the study of gravitational interaction. On the other hand, such  theories have acquired great interest in cosmology since they \lq\lq naturally" exhibit inflationary behaviours which can overcome the shortcomings of standard cosmology. From an astrophysical point of view, Extended Theories of Gravity do not require to find candidates for dark energy and dark matter at fundamental level; the approach starts from taking into account only the \lq\lq observed" ingredients ({\it i.e.}, gravity, radiation and baryonic matter); it is in full agreement with the early spirit of General Relativity but one has to relax  the strong hypothesis that gravity acts at same way at all scales. Several scalar-tensor and $f(R)$-models agree with observed cosmology, extragalactic and galactic observations and Solar System tests, and give rise to new effects capable of explaining the observed acceleration of cosmic fluid and the missing matter effect of self-gravitating structures. Despite  these preliminary results, no final model addressing all the open issues  is available at the moment, however the paradigm seems  promising in order to achieve  a complete and self-consistent theory working coherently at all interaction scales.  
\end{abstract}

\keywords{Extended Theories of Gravity; Cosmology; Quantum Field Theory; Dark Energy; Dark Matter.}

\bodymatter

\section{Introduction}
There are  serious theoretical and experimental  reasons to consider   gravity  not fully
described by Einstein's General Relativity but rather by some  alternative theory. First, attempts to renormalize General Relativity in
the $1960$s and $1970$s showed clearly that counterterms must be introduced: such terms alter the theory significantly and
make  field equations non-minimally coupled or of higher-order in derivatives  instead of second. From the physical point of view, this fact implies that
extra degrees of freedom, in addition to the usual spin two massless graviton, have  to be introduced.
Unfortunately, the corrected theory is not free of ghosts and this fact  makes it non-unitary. For example, the corrections introduced
by renormalization are quadratic in the curvature invariants and can  be adopted  in the so-called $R^2$-inflationary model for describing 
the early universe \cite{Starobinsky}. 

By retaining corrections which are generic non-linear
functions of $R$ (and no longer motivated by renormalization), one obtains the so-called $f(R)$-theories of gravity
\cite{PhysRepnostro,DeFelice,Mauro,OdintsovPR,Faraoni,Sotiriou}. Furthermore, when one tries to approach gravity (and the other interactions) from the high energy side and the weak-field limit, one does not recover exactly  Einstein's theory. For example, adopting
string theory as a full  quantum gravity approach aimed to  unify all the known interactions, one gets a low energy
limit which does not reproduce General Relativity but gives instead a scalar-tensor theory of gravity \cite{Schwarz}. Such kind of  theories has been known for a long time and were developed following initial suggestions by Dirac, Jordan, Fierz, and Thiery, culminating in  $1961$, with the paper by Brans and Dicke introducing what is now known as Brans-Dicke theory \cite{Brans}. The original motivations by  Brans
 and Dicke  were rooted in the need to implement Mach's principle, which is not fully incorporated in General Relativity,  as a 
relativistic feature  of gravity. 

After Brans-Dicke theory (the prototype of scalar-tensor theories of gravity) was
born, research on Mach's principle went its own way and, without doubt, the interest of modern physicists in
scalar-tensor gravity arises more from string and unified theories than from Mach's principle. In fact, dilaton fields and their non-minimal
couplings to the spacetime curvature are unavoidable features in any unified scheme, in particular in string theory \cite{Tseytlin}. Furthermore, first loop corrections or attempts to fully quantize gravity necessarily introduce significant deviations from General Relativity figured out as  extra degrees of freedom \cite{zerbini}. The recent spacetime thermodynamical  approach to Emergent Gravity \cite{Barcelo} pictures General Relativity as a thermodynamical state of equilibrium among a wider spectrum of gravity theories
and it  makes sense that, if extra degrees of freedom are allowed in addition to the standard massless spin two
gravitons of General Relativity, deviations from this equilibrium state will correspond to the excitation of these extra degrees of
freedom and to deviations from Einstein's theory. 
From the theoretical point of view, going beyond General Relativity is a
necessity and exploring the wider landscape of theories becomes a cultural need \cite{Faraoni}. 

From the experimental point of
view, General Relativity has been tested directly in the Solar System in its weak-field, slow motion approximations. Binary
pulsars, most notably the Hulse-Taylor system PSR $1913+16$ \cite{HT}, allow for
indirect tests outside the Solar System, in the same regime. However, strong gravity tests are still missing and
gravity is tested very poorly at the scale of galaxies and clusters, witnessing the fact that even Newtonian gravity
is doubted at galactic scales, which has led to the introduction of MOND \cite{Milgrom} and
TeVeS \cite{Bekenstein} theories to replace galactic dark matter. Cosmology cannot
be advocated as a precise test of General Relativity at largest scales: in fact, almost all theories of gravity admit the Friedmann-
Lemaitre-Robertson-Walker  solution of their field equations, with perfect fluids or other
reasonable matter sources. Indeed, it is from cosmology that comes the indication that gravity may not be
described exactly by General Relativity. 

The $1998$ discovery that the present expansion of the universe appears to be accelerated
\cite{Riess}, made using the luminosity distance versus redshift relation of type Ia
supernovae, has left cosmologists scrambling for an explanation. In order to explain the cosmic acceleration
within the context of General Relativity, one needs to introduce the mysterious dark energy, which is very exotic (its pressure P
and energy density $\rho$ must satisfy the equation $P\sim-\rho$), diluted, and comprises approximately $70-75\%$ of the energy content of the universe.  Up to now, such an ingredient has not been  detected at fundamental level. Dark energy seems  an {\it ad hoc solution} for 
the problem of  present cosmic acceleration  and alternatives have been looked for.
Attempts to explain away dark energy using the backreaction of inhomogeneities on the dynamics of the
background universe have been, so far, unconvincing. In $2002$, the idea was advanced in \cite{Capozziello}, soon followed by other authors \cite{Carroll,Nojiri}, that perhaps we are observing the first deviations from General Relativity on
the largest scales. $f(R)$-theories of gravity (although not of the quadratic form obtained by renormalization) were
resurrected in an attempt to explain this phenomenon. Curiously, $f(R)$ gravity admits a scalar-tensor
representation. Since these first attempts, the literature on $f(R)$ and scalar-tensor gravity has flourished, and these
modifications of gravity are now proposed as reliable alternatives to dark matter and dark energy. It would be
premature and unjustified to claim that gravity is described by any of these theories. However, even if none of
these Extended Theories of Gravity ultimately proves to be correct, they could solve many problems while still
allowing us to peek into the vast landscape of theories beyond General Relativity and to understand many ways in which gravity
can be enlarged with respect to Einstein's conception \cite{Faraoni}. Here, with no claim to be complete, we outline some research topics and issues that could be addressed and framed within the realm of Extended Gravity.

\section{Cosmological and astrophysical motivations: dark energy and dark matter}

Recently, data coming from astrophysical and cosmological observations result of extremely high quality and, globally taken, lead to the so-called Precision Cosmology. This new
era of research is bringing to extend accurate methods of investigations, appropriate for experimental physics,
also to cosmology and, generally, to astrophysics. The emerging picture gives a representation of the universe substantially new with respect to the Standard Cosmological Model, assumed as a paradigm up to the beginnings of nineties.
Such a model was essentially based on General Relativity, the standard theory of elementary
particles, the initial Big Bang and a succeeding phase of decelerated expansion in which the cosmos evolved
through a radiation dominated phase and a following matter dominated one, in which structures have been
formed. Inflationary Theory, formulated by means of various approaches during eighties, seemed to have
ruled out several incongruences of this model, having given plausible mechanisms to remove initial
singularity, to explain large-scale structure formation and to explain, within a coherent theoretical
framework, the (observed) absence of topological defects like magnetic monopoles and cosmic strings. From
Precision Cosmology, instead, a picture emerges in some way surprising, according to which also the
relatively near (at small redshifts) universe results very different as to the representation we had of it till
nineties.

 In synthesis, the universe can be represented as a spatially flat manifold with an ordinary matter-energy content ({\it i.e.} baryonic matter and radiation) well below the critical value necessary to obtain flatness from the Einstein-Friedmann equations. 
 Furthermore, cosmological standard candles, used as distance indicators, suggest an accelerated expansion phase, hardly obtainable once we consider ordinary fluids as the source of cosmological equations \cite{Peebles,Padmanabhan,Padmanabhan2}. 
 
 This latter evidence has been a true surprise that created
difficulties also for explaining the genesis and evolution of the observed large scale structures \cite{Peebles2,Peacock}.
 
 The results of several observational campaigns have contributed to this new picture. They are, essentially, the estimates of galaxy
cluster masses \cite{Allen}, the correlation functions \cite{Younger} and the numerical cluster abundances in terms of redshift \cite{Wang}, the Hubble diagram derived from type-Ia supernovae (SNeIa) observations \cite{Riess}, the optical
surveys of large scale structure \cite{Cole}, the measurements of cosmic microwave
background radiation (CMBR) anisotropies \cite{Spergel}, the measurements of
cosmic shear through gravitational weak lensing surveys \cite{Schmidt} and, finally, data
on Lyman alpha forest absorption lines \cite{McDonald}. It is realistic to say that
the interpretation of this huge and increasing amount of information, within the same unitary and self-consistent
theoretical framework, constitutes the biggest challenge of modern cosmology. Specifically, the existing
discrepancy between the observed luminous matter and the critical density, needed to obtain a spatially flat
universe and then to give rise to the accelerated expansion, can be only filled if one admits the existence of a
cosmic fluid, with negative pressure, which does not result clustered as in the large scale structure. In the simplest
scenario, this mysterious ingredient, known as dark energy, can be represented as the Einstein cosmological
constant $\Lambda$ and would account for about $70\%$ to the global energy budget of the Universe. The remaining $30\%$,
instead clustered in galaxies and clusters of galaxies, should be constituted for about $4\%$ by baryons and for the
rest by cold dark matter (CDM), theoretically
describable through WIMPs (Weak Interacting Massive Particles) or axions.

From a genuine astrophysical point of view, this simple model has the
appreciable feature to be in very good agreement with observations. It can be reasonably assumed as the first step
towards a new standard cosmological model, the Concordance $\Lambda$CDM Model \cite{Bahcall}.
In synthesis, the
presently observed universe could be coherently described once we admit the presence of a cosmological
constant ($70\%$ of the total content), which would give rise to the observed acceleration of the Hubble fluid,
and the presence of dark matter (at least $25\%$), which would explain the large scale structure.

 Notwithstanding the satisfying agreement with observations, the $\Lambda$CDM model presents several
incongruences and shortcomings. If the cosmological constant constitutes the {\it vacuum state} of the gravitational
field, we have to explain the $120$ orders of magnitude between the observed value at cosmological level and the
one predicted by any quantum gravity theory \cite{Weinberg,guendelman}. Furthermore, there is the
so-called {\it coincidence problem}, for which the matter (both dark and baryonic) and the cosmological constant
(dark energy) are today of the same order of magnitude (being, for the cosmological evolution, $30\%$ and $70\%$
very similar numbers), even with completely different dynamics, according to the cosmological equations.

Several models have been proposed to address these issues. However, none of them is fully satisfactory, from
both theoretical and observational points of view, first of all because none of the suggested candidates for dark
matter and dark energy has been experimentally detected.

 Secondly, the $\Lambda$CDM model is not able to fully explain
several observational evidences at scales of galaxies and galaxy clusters, that is, it presents some incongruences
at scales smaller than cosmological ones. If, from one hand, the flat rotation curves, observed since seventies in
spiral galaxies, have reinforced the old hypothesis of the necessity of dark matter (that has thus become an
irreplaceable ingredient in the $\Lambda$CDM model), the low concentration of the dark haloes observed in some systems
(like low surface brightness (LSB) galaxies) appears to be in disagreement with the predictions coming from N-body
cosmological simulations for the $\Lambda$CDM model \cite{deBlock}.  
Analogously, the
existence of elliptical and S0 galaxies, with an extremely variable dark matter amount, could be due to both a
different efficiency of the star formation processes and a manifestation of the same concentration effect found in
LSB galaxies, substantially in disagreement with the $\Lambda$CDM model \cite{Napolitano1}.

Also, the presence of a central cusp in the dark matter distribution of galaxy clusters, predicted by theoretical
models and simulations, has not been determined with certainty \cite{Cardone}. Up to now, it is not clear whether such discrepancies are due to observational problems, to the lack of
understanding the mechanisms that rule the baryon physics (for example, in N-body simulations used to predict
properties of dark matter distribution), or to a substantial failure of the $\Lambda$CDM model.

We have to notice that, in the history of science, similar situations emerged several times and, in order to
sustain or disprove a model, new ad hoc ingredients have been introduced, like \lq\lq epicycles" and \lq\lq deferents"
in the ancient geocentric models or, more recently, like ether for explaining electromagnetic waves within
the realm of classical physics. A paradigmatic example can be that of the discovery of Neptune by Galle:
the perturbed Uranus orbit could only be explained, within Newton theory, admitting the existence of another
planet, as predicted by Le Verrier and Adams.

Turning back to the  $\Lambda$CDM model, the paths are essentially two: either we go on looking for the dark components till we do not actually find them (like in the Neptune case)
or we admit that the cosmic acceleration and the \lq\lq missing" mass are nothing else but signals that General Relativity, actually
tested in the range of laboratory and Solar System scales, is unable to describe the universe at larger scales (in such a case, dark matter and dark
energy would have a role like that of ether, which became useless with the advent of a new self-consistent
theory as Special Relativity).

In
this case, extensions of General Relativity, like $f(R)$-gravity, should be invoked \cite{CapFra,Capozziello1}. In order to be fully satisfactory, Extended Gravity should be accurately tested at all scales. Considering critically the approach, we have an almost equivalent description with respect to dark energy and dark matter Universe and then the turning point would be to find out some \lq\lq experimentum crucis" capable of discriminating between the two pictures. From an astrophysical viewpoint, we have at our disposal an adequate amount of observational data related to several mass tracers in galaxies and
clusters of galaxies, and it is now conceivable to individuate experiments suited to verify or falsify the Extended
Gravity approach. In fact, recent technological developments have remarkably increased the possibility to study
properties of mass distribution (radial profile and shape) around a single galaxy or a cluster of galaxies, using
long range tracers such as planetary nebulae \cite{Napolitano}, globular clusters \cite{Richtler}, HI disks in spiral and polar ring galaxies \cite{Arnaboldi} and diffused $X$-ray emission \cite{OSullivan}, analysis of rotation curves \cite{salucci}
or techniques like
gravitational lensing \cite{Koopmans}.  On the other side, it is necessary to look for correlations among the
investigated quantities in order to frame some fundamental empirical relationships, such as the Tully-Fisher relation,
within Extended Theories of Gravity \cite{Troisi,Mendoza}. In this way, such relations would assume an intrinsic meaning,
that is, without the necessity to justify them with the aid of dark matter. Besides, for extending this kind of
investigation to elliptic galaxies, too, velocity dispersion can be considered instead of rotation curves. Specifically, using the gravitational potentials
induced by Extended Theory of Gravity, the reconstructed mass profile should be a \lq\lq tracer" of the optical luminosity, and the rotation curves giving evidence of the absence of dark matter. 
Furthermore, in particular, stellar systems are an
ideal laboratory to look for signatures of possible modifications
of standard law of gravity. In fact, differences
stemming from the functional form of $f(R)$ or  scalar-tensor theories may prevent
the existence of relativistic stars in these theories \cite{Briscese,Nojiri2}.
However, possible problems jeopardizing the existence
of these objects may be avoided \cite{Hu, Tsujikawa,Upadhye} by considerations
on the chameleon mechanism \cite{Khoury}.
 
In particular, the Chameleon mechanism, firstly adopted  in scalar-tensor gravity,   can effectively reduce locally the non-minimal coupling between the scalar field and matter. This mechanism is invoked to reconcile large-scale departures from General Relativity, supposedly accounting for cosmic acceleration, to small scales stringent constraints on General Relativity.
 It was investigated this framework on cosmological and Solar System scales to derive combined constraints on model parameters, notably by performing a non-ambiguous derivation of observables like luminosity distance and local post-newtonian (PPN) parameters. Likelihood analysis of type Ia Supernovae data and of admissible domain for the PPN parameters clearly demonstrates that chameleon mechanism cannot occur in the same region of parameters space than the one necessary to account for cosmic acceleration with the assumed Ratra-Peebles potential and exponential coupling function \cite{Fuzfa}.

On the other hand,
some observed stellar systems are incompatible with the
standard models of stellar structure \cite{Babichev}: these are peculiar
objects, as star in instability strips, protostars or
anomalous neutron stars (the so-called \lq\lq magnetars" \cite{Muno}
with masses larger than their expected Volkoff mass) that
 could admit dynamics in agreement with modified gravity
and not consistent with standard General Relativity. It seems that, on
particular length scales, the gravitational force is larger
or smaller than the corresponding General Relativity value. For example,
the addition of $R^2$ terms, in the Hilbert-Einstein
Lagrangian, allows a major attraction while a $R_{\alpha\beta}R^{\alpha\beta}$
term gives a repulsive contribution \cite{Arturo}. The corrections are essentially Yukawa-like and the physical implication is that new characteristic
lengths emerge beside the standard Schwarzschild radius related to the gravitational mass. This feature can have dramatic
consequences in astrophysics and cosmology. In conclusion, the Extended Theories of Gravity or any theoretical scheme
with the aim to explain astrophysical structures or global dynamics of the universe can be tested or
contradicted only through a strict comparison with observations \cite{annalen,idro,jeans}.

\section{Inhomogeneous cosmologies}

A key role to test Extended Gravity could come from inhomogeneous cosmologies. For a precision estimate of
the dynamical influence of the inhomogeneities on the large-scale evolution of cosmological backgrounds, a
covariant and gauge invariant averaging procedure, valid for both space-like \cite{Gasperini,Gasperini1} and null \cite{Gasperini2} hypersurfaces can be
developed. Such an (exact) procedure can be applied to obtain a full computation, to the second perturbative
order, of the luminosity-redshift relation in a CDM \cite{Ben} and $\Lambda$CDM dominated
backgrounds, perturbed by inhomogeneity fluctuations of primordial (inflationary) origin. Preliminary
results have shown that the contribution of \lq\lq realistic'' geometric fluctuations induces, in general, non-negligible
backreaction effects on the averaged luminosity-redshift relation, but such effects seem to be too small to mimic a
sizable fraction of dark energy. However, the dispersion due to stochastic fluctuations is much larger than the
backreaction itself, implying an irreducible scatter of the data that may limit to the percent level the precision
attainable on cosmological parameters (at least in the context of current astronomical observations). In particular, several new solutions in modified gravity, in particular in $f(R)$-gravity, have been found analytically and checked numerically \cite{Ivanov}. This is a
fundamental issue to retain or rule out any alternative theory of gravity with respect to dark components.

\section{Relic abundances}

The new physics beyond the Standard Model of interactions, which predicts the
existence of particles which are candidate for dark matter constituent, is nowadays under deep scrutiny. From a
side, accelerators (like the LHC at the CERN) will provide informations aimed to test theories of new physics
beyond the Standard Model, from the other side, cosmology and astroparticle physics offer a unique scenario
for determining the basic properties of the dark matter candidates. In this framework, the supersymmetric
particles, such as WIMPs (weakly interacting massive particles) could be the leading candidate to explain the
nature of dark matter since their thermal production in the early universe may give rise to a relic density of the
same order of magnitude of the present dark matter density. Particularly relevant in these scenarios could be the
role played by Extended Theories of Gravity. This is due to the fact that these theories predict a thermal
evolution of the universe different with respect to the one based on General Relativity. More precisely, Extended Gravity predicts
a modification (e.g. amplification) of the expansion rate of the universe with respect to the standard cosmology so
that the thermal relics decouple with larger relic abundances. As a consequence, the correct value of the relic
abundance comes out from larger annihilation cross-section. An immediate applications of these alternative
cosmologies is to provide the correct value of the cross-section of thermal relics able to explain the recent data
of PAMELA experiment \cite{Catena,pamela}.

\section{Mixing fields, vacuum fluctuations and dark energy}

It has been recently shown that a close connection
between the mixing phenomenon and dark energy problem exists: many evidences lead to think that this
mechanism could give an explanation to the cosmic acceleration in the realm of quantum physics \cite{Capolupo}. Furthermore classical and quantum fluctuations can be described in terms of marginal distribution functions in the framework of tomographic cosmology \cite{cosimo}. 
Actually, the experimental evidences connected to neutrino oscillations are one of the
most important discoveries of today particle physics, and, consequently, theoretical studies of particle mixing
(quarks, neutrinos, mesons) and oscillation phenomena have intensified more and more. For example, it has
been shown that the vacuum for mixing fields (flavour vacuum) has the structure of a condensate of particle-antiparticle
couples, for both fermions and bosons, so that observable quantities, like oscillation effects, present
corrections which could result extremely interesting for explaining the cosmological fluid giving rise to
acceleration. On the other hand, although we are still far from having a clear idea of what a full-fledged
theory of Quantum Gravity would eventually look like, semi-classical approaches to this issue led to the seminal
idea that quantum vacuum fluctuations, considered as a form of energy, must \lq\lq gravitate", that is, must enter into
the vacuum expectation value of the stress-energy tensor \cite{Zeldovich}. The key
question is: do quantum vacuum fluctuations fullfil the equivalence principle of Relativistic Theories of Gravity?
This seems to have been clearly settled down \cite{Fulling}, but there are still
contradictory answers which pop up in the literature. In the
laboratory, quantum fluctuations have been proven to exist, without reasonable doubt. 
Although some (important) claims have been risen that
vacuum contributions cannot be isolated and that they do not have an absolute meaning in terrestrial labs
(since they do not couple with any quantum field, aside from gravity) they do appear in the energy
momentum tensor of General Relativity, and their effects are quite noticeable at the nanoscale, as manifested in the dressing
of quantum particles, in the Lamb shift effect (recently reported in Science to have even been observed in
solid-state superconducting systems \cite{Fragner}), and specially in energy-difference calculations in a number of
different cases of the Casimir effect and its extensions, notably the Casimir-Lifschitz theory. Depending on
the specific configuration, vacuum fluctuations can give rise to attractive or repulsive forces \cite{Elizalde} and some
laboratory proofs  of the existence of the later has been reported \cite{Munday}.
This type of effects could
be the source of cosmic acceleration through quantum vacuum fluctuations: the final result could be an effective
probe consistent with the Extended Gravity approach.

\section{Baryon/Leptogenesis}

In the framework of the Extended Gravity, the origin of  matter-antimatter
asymmetry  and  Leptogenesis are other fundamental issues. These studies can be
based on the coupling between the (Majorana) neutrino and the gravitational background. As well known, the
matter-antimatter asymmetry in the universe is still an open problem of the particle physics and cosmology.
The successful prediction of Big-Bang nucleosynthesis demands on the assumption that the net baryon number
to entropy ratio must be of the order $10^{-10}$, which is compatible with WMAP data on CMBR. As shown by
Sakharov, a (CPT invariant) theory is able to explain the baryon asymmetry provided that the following
conditions are fulfilled: 
\begin{enumerate}
\item
there must exist processes that violate the baryon number; 
\item the discrete symmetries
C and CP must be violated; 
\item  departure from thermal equilibrium.
\end{enumerate}
 However, these conditions may be relaxed in
some circumstances.
 A dynamical violation of
CPT (which implies a different spectrum of particles and antiparticles) may give rise to the baryon number
asymmetry also in a regime of thermal equilibrium \cite{Choen}. Moreover, a successful mechanism for explaining the
asymmetry between matter and anti-matter is provided by the Leptogenesis \cite{Fukugita}. In this scenario, where the Majorana neutrino is introduced, it is possible to generate the
baryon asymmetry if the asymmetry is generated in the lepton sector at either Grand Unified Theory (GUT) or
intermediate scales, even if the baryon number is conserved at high energy scales. Sphaleron processes convert
the lepton asymmetry in the observed baryon asymmetry. The GUT framework embeds quite naturally the see-saw
mechanism, which requires the existence of very heavy right-handed Majorana neutrinos in order to explain
the neutrino mass suppression. Is it possible to investigate a scenario for baryogenesis, via leptogenesis, in a
$SO(10)$ inspired model. Exploiting recent experimental results from Daya Bay and RENO, the right handed
neutrino spectrum can be predicted, and, most importantly, both the size and the sign of the Baryon Asymmetry
of the universe (BAU). Moreover, the role of cosmological scenarios provided by Extended Gravity in the
Baryo/Leptogensis problem can be investigated by means of coupling of Majorana neutrinos with the
gravitational background.

\section{CP violation and Cabibbo-Kobayashi-Maskawa (CKM) mechanism}

Careful analysis of new available
experimental data seems to be necessary to unveil the origin of CP violation. In fact, CP violation in Bs hadronic
decays have been  measured at LHCb\cite{Fleischer,Fleischer1}, that is $ Bs\rightarrow J/\Psi \eta'$ and $Bs\rightarrow J/\Psi f0(980)$. These decays are also an interesting
probes for the determination of the $SU(3)$ octet/singlet mixing parameters related to the pseudo scalar $\eta'$ and
the (possibly exotic) $f0(980)$, respectively. Further phenomenological studies of mixing in $\eta-\eta'$ have been
presented in \cite{didonato}. The determination of CKM parameters is
important for assessing the size of CP violation in the Standard Model (SM) and beyond. As of today, there are
a few discrepancy between theory and experiment in the extraction of the CKM parameter $ |Vub|$, that have been
investigating in several papers  \cite{Aglietti}. This
phenomenon could be a very important signature for Extended Gravity at fundamental level.

\section{Neutrino oscillations, gravitational waves and neutrino mass absolute value}

The determination
of the absolute values of neutrino masses is certainly one of the most difficult problem from the experimental
point of view. One of the main difficulties in determining the neutrino masses from Solar or
atmospheric experiments concerns the ability of neutrino detectors to be sensitive to the species mass-square
difference instead of the neutrino mass itself. A model-independent method to achieve this goal has been
recently introduced \cite{Lambiase}. It is possible to show that the
neutrino mass-scale can be directly achieved by measurements of the delay in time-of-flight between the
neutrinos themselves and the gravitational waves burst generated by the asymmetric flux of neutrinos
undergoing coherent helicity (spin-flip) transitions during either the neutronization phase, or the relaxation
(diffusion) phase in the core of a type II SNae explosion. Because special relativistic effects do preclude massive
particles of traveling at the speed of light, while massless do not (the standard graviton in this case), the
measurement of this neutrinos time lag leads to a direct accounting of its mass. Then the accurate measurement
of this time-of-flight delay by SNEWS + LIGO, VIRGO, BBO, DECIGO experiments might readily assess the
absolute neutrino mass spectrum. In this framework, a lower bound on the neutrino mass can be obtained. In
fact, assuming that the time delay between the gravitational waves and neutrino burst (with energy 10 MeV) is
10 msec and that neutrino sources are at distances of 2 Mpc, the value of 0.01eV, compatible with the present
estimation on neutrino mass, can be obtained. These estimations could be a very important tool to probe
Extended Gravity  as background for these phenomena \cite{gae}.

\section{Further gravitational modes and quadrupolar radiation}
Also the post-Minkowskian limit of Extended Gravity  deserves an accurate consideration, in particular with respect to the problem of gravitational radiation. In fact, further possible \lq\lq signatures" could come from gravitational radiation. It
has pointed out that Extended Gravity gives rise to new polarizations for the gravitational waves with respect
to the standard + and $\times$ polarizations of General Relativity \cite{PLB}. In particular, one find that besides a massless spin-2 field (the standard graviton), the theory contains also
spin-0, spin-2 massive and ghost modes.  Then, it could be possible, in principle,  to investigate the detectability of such additional
polarization modes  by ground-based and space interferometric detectors, extending the formalism of the cross-correlation analysis, including the additional polarization
modes. In particular, it could be possible  to detect the  energy spectrum density of the   stochastic background of
 relic gravity waves \cite{greci}.
This seems a very
significant feature for several relativistic theories of gravity that could be investigated by ground based and
space interferometric experiments \cite{Bellucci,greci}.
Furthermore, the debate concerning the viability of $f(R)$-gravity as a natural extension of General Relativity could be
realistically addressed by using results coming from binary pulsars like PSR $1913 + 16$ and from other binary systems. To this end, 
 a quadrupolar approach to the gravitational radiation can be developed for  analytic $f(R)$-models. It is possible to show that experimental results are compatible with a consistent range of $f(R)$-models. This means that
$f(R)$-gravity is not ruled out by the observations and gravitational radiation (in strong field regime) could
be a test-bed for such theories \cite{quadru}. This result implies a revision of the gravitational wave physics  and  the eventual detection of gravitational massive modes (or, in general,  modes
different from those predicted by General Relativity) could be the final \lq\lq experimentum crucis" for Extended Gravity.

\section{Black holes, wormholes,  and high-energy gravitational scattering }

In all areas of physics and mathematics it is common to search for insight into a theory by finding exact solutions of its
fundamental equations and by studying these solutions in detail. This goal is particularly difficult in non-linear theories and
the usual approach consists of assuming particular symmetries and searching for solutions with these symmetries. Stripped
of inessential features and simplified in this way, the search for exact solutions becomes easier. In a sense, this approach
betrays a reductionist point of view but, pragmatically, it is often crucial to gain an understanding of the theory that cannot
be obtained otherwise and that no physicist or mathematician would want to renounce to.
It is therefore quite natural to ask about black hole features in those gravitational theories since, on the one hand, some black holes signatures may be peculiar to Einstein's gravity and others may be robust features of all generally covariant theories of gravity \cite{zerbini2}. On the other hand, the obtained results may lead to rule out some models which will be in disagreement with expected physical results. The study of the structure of these black objects, in Extended Theories of Gravity, is interesting for a wide variety of reasons. 
Although black holes are one of the most striking predictions of General Relativity, they remain one of its least tested concepts. Electromagnetic observations have allowed us to infer their existence, but direct evidence of their non-linear gravitational structure remains elusive. In the next decade, data from very long-baseline interferometry  and gravitational wave detectors should allow us to image and study black holes in detail. Such observations will test General Relativity in the dynamical, non-linear or strong-field regime, precisely where tests are currently lacking.
Testing strong-field gravity features of General Relativity is of utmost importance to physics and astrophysics as a whole \cite{gerardo}. This is because the particular form of black holes solutions, such as the Schwarzschild and Kerr metrics, enter many calculations, including accretion disk structure, gravitational lensing, cosmology and gravitational waves theory. The discovery that these metric solutions do not accurately represent real black holes could indicate a strong-field departure from General Relativity with deep implications to fundamental theory. 
Such tests require, for example, parametrizing deviations from Schwarzschild or Kerr solutions \cite{KerrETG,S,Tsukamoto}. 
It is interesting to see that while spherical solutions are essentially
Schwarzschild-like in General Relativity (with constant or null curvature), in $f(R)$-gravity it is possible to have several possibilities essentially related
to the fact that also curvature dependent on radial coordinate gives rise to spherically-symmetric solutions \cite{stabile}. Furthermore,  it is possible to generate axially symmetric solutions, starting from
spherical ones, by the Newman - Janis algorithm \cite{KerrETG}.

In particular, in $f(R)$ extensions of General Relativity, the Palatini approach provides ghost-free theories with second-order field equations that allows to obtain charged black hole solutions which depart from the standard Reissner-Nordstr\"om solution \cite{Rubiera}. 

A particular mention deserves the search for wormhole solutions. Such exotic structures could constitute a formidable  signature for Extended Gravity and represent an extremely interesting \lq\lq lab" to test physics at fundamental level \cite{harko}. 

 Furthermore, the problem of
understanding whether and how information is preserved in a quantum process whose classical analog leads to
black hole formation has been addressed for a number of years in the context of superstring theory, supposedly
a consistent framework for combining General Relativity and quantum mechanics. In one approach, one considers the
transplanckian energy collision of two massless strings as a function of center of mass energy and impact
parameter \cite{Amati,Veneziano,Ciafaloni}, in another approach \cite{DAppollonio} one looks instead at the
high energy collision of a massless closed string on a stack of N D-branes, again as a function on energy and
impact parameter, but also of N (since this enters in the characteristic scale of the effective geometry generated
by the branes). Much progress has been made, in both cases, about reproducing, from first principles, the effects
of an emerging geometry (such as gravitational deflection or tidal excitations), but the most interesting regimes
(the one in which a black hole is expected to be formed in the first case, or the closed string should be absorbed
by the brane system, in the second) are still far  to be understood. The ultimate goal would be to understand how
information about the pure initial state gets encoded in the complicated final one which, in many respects,
should look like a mixed, thermal state, and yet carry no entropy. These results would be extremely important to
discriminate among effective theories of gravity, possibly different from General Relativity, since higher energy corrections in
the scattering processes give contribution to the interaction Lagrangian as high curvature terms.

\section{Conclusions}
In conclusion, the Extended Theories of Gravity (or any theoretical scheme with the aim to explain
astrophysical structures and the global dynamics of the universe) can be tested or ruled out by means of a strict
comparison with observations. Retaining or rejecting a given picture strictly depends, on one side, by finding
out self-consistent signatures at fundamental level and, from the other side, by consistently fitting the most
possible amount of observational data at several different scales. In principle, the correct interplay between features at  IR and UV scales should lead us to discriminate among  concurrent  theories of gravity.  

This approach poses interesting
problems related to the strict validity of General Relativity. Such a theory
works very well at local scales (Solar System) where effects of further
gravitational degrees of freedom are hard to be  detected with present day facilities \cite{vecchiato}.  As soon
as one is investigating larger scales, as those of galaxies, clusters of
galaxies, etc., further corrections can be introduced in order to explain
both astrophysical large-scale dynamics  and cosmic evolution. Alternatively, huge amounts of dark matter and dark
energy have to be invoked to explain the phenomenology, but, up
today there are no final evidences for these new constituents at
fundamental level. This lack seems related, in a very subtle way, to the Quantum Gravity issue \cite{modesto}. A further approach can be that to lead a deep analysis of further degrees of freedom coming from Extended Gravity. In particular, General Relativity can be assumed in its metric formulation while further degrees of freedom, coming for example from $f(R)$-gravity, in the Palatini metric-affine formalism \cite{harko1,harkocap}. This hybrid approach could avoid inconsistencies and shortcomings of both metric and metric-affine approaches, singly considered, and allow to bypass problems that emerge in  astrophysical and cosmological contexts \cite{viriale}.  However, this new formulation needs to be deeply and critically investigated. 

As final remark, we can say that Extended Gravity seems a fruitful paradigm to investigate self-gravitating structures ranging from stars up to the whole universe preserving the robust result of General Relativity achieved at local and Solar System scales. Despite of this positive feature, today we are far from a final, self-consistent theory embracing all the gravitational phenomenology from quantum to cosmic scales.

\end{document}